# VALIDATED REAL TIME MIDDLE WARE FOR DISTRIBUTED CYBER PHYSICAL SYSTEMS USING HMM

Ankit Mundra[1], Bhagvan Krishna Gupta[2], Geetanjali Rathee[3], Meenu[4], Nitin Rakesh[5] and Vipin Tyagi[6]

[1,2,3,4,5]Department of Computer Science Engineering, Jaypee University of Information Technology, Waknaghat, Himachal Pradersh, India

[1]ankitmundra8891@gmail.com ,
[2]bkgupta21@gmail.com,
[3]geetanjali.rathee123@gmail.com
[4]meenuchawla011@gmail.com,
[5]nitin.rakesh@gmail.com

[6]Department of Computer Science and Engineering, Jaypee University of Engineering and Technology, Guna, Madhya Pradesh, India

[6]dr.vipin.tyagi@gmail.com

*ABSTRACT*

*Distributed Cyber Physical Systems designed for different scenario must be capable enough to perform in an efficient manner in every situation. Earlier approaches, such as CORBA, has performed but with different time constraints. Therefore, there was the need to design reconfigurable, robust, validated and consistent real time middle ware systems with end-to-end timing. In the DCPS-HMM we have proposed the processor efficiency and data validation which may proof crucial in implementing various distributed systems such as credit card systems or file transfer through network.*

## 1. INTRODUCTION

Distributed systems are the systems which deal with hardware and software systems containing more than one processing element or storage element, concurrent processes, or multiple programs, running under a loosely tightly controlled regime. Special care must also be taken that messages (e.g., Transaction in credit card system, intermediate results in medical field) are indeed delivered correctly and that invalid messages (e.g., false transaction, and incorrect/insufficient results) which would otherwise bring down the system efficiency and perhaps the rest of the network, are rejected. Cyber physical systems are the need of today and hence they must be made compatible enough to handle various types of events may be periodic or aperiodic whose validation is critical to the correct behavior of the system. e.g., in an industrial plant monitoring system, an aperiodic alert may be generated when a series of periodic sensors observations/readings meet certain detection criteria [1]. Varying user inputs and certain other parameters may trigger other real-time aperiodic events.

While traditional real time middle ware systems such as real time CORBA [2] and Real time Java [3] have shown promise as distributed software platforms for systems with time constraints, existing middle ware systems lack the flexibility and validation of the results/outputs. e.g., planned allocation is an effective mechanism for handling variable real time workloads in Distributed Cyber Physical Systems. While this allocation may not be applicable in any other system so these algorithm needs to be more specific with respect to the CPS. However, job swapping can be applied to improve the processing of the system and thus reducing the complexity. But this cannot be applied in critical control systems in which certain delay in processing may cause sudden disaster on the system. Therefore a key challenge is to design a Distributed Cyber Physical System which is flexible, efficient and can also validate the results. The system should be able to support the diverse requirements of Distributed CPS.

Providing middle ware services with configurable strategies faces: (1) the specific criteria which decides the nature of the task and allocates the resources accordingly. (2) Appropriate combination of the available resources. (3) Suitable validation criteria so as to make the system secure from the false outputs thus making the system robust. To deal with these challenges we have designed and implemented a new set of component middle ware services including end-to-end process/sub process scheduling, resource allocator and validation criteria. We have developed the front-end implementing the algorithm in order to test the algorithm.

On the other hand Cyber physical system is the new generation architecture which incorporates computation and communication with physical control process. Now a day's CPS is viewed as a new science for future engineered. It is a system with controlled physical integrated communication and computational capabilities. Because of advancement in modern technology where physical interaction needs cyber physical system gets the attention of researcher and can be viewed for both developments in academic and industrial field.

In this paper, we have proposed a middle ware system which works on the nature of the process thus, provides key features such as flexibility (as applicable on different scenarios), efficiency, security and Robustness. The proposed DCPS-HMM provides different modules for management, allocation of resources and the output are passed through the validation module which increases the credibility of the proposed cyber physical system.

This paper is organized in five sections. Section first is the brief introduction to validated real time middle ware for distributed cyber physical systems. Section second presents related and previous work for validation of middleware. Further section third details the DCPS-HMM model, the component overview, working of the model and characteristics of DCPS-HMM are stated. Section fourth describes the simulation of proposed model in the two scenarios one is credit card fraud detection and second is CPS-IP scenario using DCPS-HMM Model. The research paper is concluded in section fifth and the future work in this field is mentioned in this section.

## 2. RELATED WORK

The Validation Criteria for checking the consistency of data has drawn a lot of research interest and developed number of techniques using data mining and neural networks. S. Ghosh et. al., have proposed credit card fraud detection with neural network [4]. They have built a detection system, which is trained on a large sample of labeled credit card account transaction. These transactions contained example fraud cases due to lost cards, non- received issue and many more.

M. Syeda et. al., have used parallel granular neural networks (PGNNs) for improving the speed of data mining and knowledge discovery process in credit card fraud detection [6]. They have built a detection system, which is trained on large sample of earlier observations. The major problem with the earlier approaches, [7], [8], [9], and [10] is that they require labeled data for genuine inputs as well as for the false data. These approaches do not detect the false data where labeled data is not available.

Object Management Group provides a standard to open distributed object computing infrastructure known as Common Object Request Broker Architecture (CORBA) [1]. CORBA supports the network management tasks i.e. registration of objects; tracking object location; multiplexing and de-multiplexing; request packet framing and error-handling; marshaling and un-marshaling of variables and parameter. But again this system is not dealing with flexibility and validation criteria.

In contrast, we have proposed a Validation criteria PV, which do not require any labeled data and yet able to determine the validity of the results by following the past track of the system's results and initial data, e.g., credit card fraud detection is characterized by the previous spending habits of the credit card holder [5].

## 3. DCPS-HMM MODEL

To support end-to-end periodic and aperiodic events in distributed cyber physical systems, we have proposed architecture for process management and process validation. The key feature of this approach is the validation of results for different kinds of periodic and aperiodic events. Our framework provides PA, PI, PT and PV which interact with the application through PM. As a part of DCPS-HMM model, we consider CPS comprises of physical systems generating aperiodic and periodic [1] events. Hence the set of the sequence of events is referred to as a process. A process is further composed of sub processes $P_{i,j}$ where $j$ varies from $1$ to number of sub processes located on $n$ different processors. The release of execution of a sub process is preceded by the execution of the previous sub process and is followed by subsequent execution of the sub processes following it. Periodic events are those events which are compatible with the phenomenon of job swapping; whereas aperiodic events are the ones which must be executed in one go. The proposed system consists of following components:

### 3.1. Component Overview

The proposed system consists of following components:

#### 3.1.1. Process Manager (PM)

All other component of the system interacts with the application components through process manager. The PM provides a process assignment to the processor allocator. The assignment of various processes is taken care by a queue. Each time the allocator allocates the processor to a process or a sub process and is de-queued from the PM. If there is no process left then PM will become idle and will wait for further user interaction. It also receives the report from the process tracker about the completion of all the sub tasks on one processor. Therefore it also keeps tracks of the processor status which is updated in the processor allocator thus avoiding an overloading of one processor.

### 3.1.2. Process Allocator (PA)

The foremost function of the processor allocator is to characterize the process / sub processes to their nature and thus program the processor allocated accordingly. Thus it keeps the track of the processor present state. If the periodic process is being executed then the criteria of the job swapping algorithm can be applied in order to improve the efficiency of the system. Otherwise, other criterion could be applied, under consideration that process is executed in one go rather than in chunks.

### 3.1.3. Process Implementation (PI)

This component performs the desired operation on the system which is required by the process. E.g., in credit card scenario the transaction is performed and then further validated by the process validator [4].

### 3.1.4. Process Tracking (PT)

The foremost function of the tracker is to report all the completed sub tasks on one processor to the PA component when the processor becomes idle, so that PA can remove their expected utilization to reduce the worst case scenario of the system.

### 3.1.5. Process Validator (PV)

The PV is double embedded process with two hierarchy levels. This component works on finite set of states determined by a set of transition probabilities. The state determined by the algorithm is not visible to an external observer. The basic idea behind this model/component is to simulate the behavior of the given system. A PV can be characterized by the following [5].

N is the number of the states in the component. Set of states are denoted as $S = \{S_1, S_2, \ldots S_N\}$, where i=1,2,3,............N is an individual state. The state at time $t$ is denoted by $E_t$.

M is the number of distinct observations symbols per state. The observation symbol corresponds to the physical output of the system being simulated. We denote the set of symbols $R = \{R_1, R_2, \ldots, R_M\}$, where $R_j = 1,2,3,4,\ldots, M$ is an individual symbol.

The state transition probability matrix $T1 = [t_{1_{ij}}]$,

where $t_{1_{ij}} = P(E_t + 1 = S_j | E_t = S_i)$.

For the general case $j$ can be reached from any other state $i$ in a single step.

Then we define the observation symbol probability matrix $T2 = [t_{2_j}(k)]$.

where $t_{2_j}(k) = P(R_k = S_j)$.

The initial state probability vector $w$ where $w_i = P(E_i = S_i)$.

The observation sequence $O = O_1, O_2 \ldots O_R$. An observation sequence can be generated by many possible state sequences. $E = E_1, E_2 \ldots E_R$, where $E_1$ is the initial state.

Further, we consider three probability distribution T1, T2 and *w*. We use the notation C to indicate the complete set $C = (T1, T2, w)$.

The probability generated from this state sequence is given by:

$P(O|E, C) = \prod_{t=1}^{M} P(O_t|E_t, C)$, where statistical independence of observation is assumed.

We map the actual data into M ranges $R_1, R_2, \ldots, R_M$ forming the observation symbols.

The actual range is defined for each symbol is configurable based on the previous data available to the CPS, e.g., In credit card scenario let's say we consider three price ranges namely low(*l*), high(*h*) and medium(*m*) [5]. Therefore our set of observation is $R = \{l, m, h\}$ making M=3. This set of observation symbols are formed dynamically.

For this we use K-means clustering technique to determine the clusters [4]. The number of clusters K is fixed a priori. The grouping is performed by minimizing the sum of squares of distances between each data point and the centroid of the cluster to which it belongs. Here K is same as the number of observation symbols M. Centroids are generated by the clusters and are used to decide the new observation symbol when any new data comes in.

Now, considering the second layer where we start with an initial estimate of PV parameters A, B, initial state probability vector. Initial state probability distribution is considered to be uniform, that is, if there are N states then the initial probability of each state is 1/N. This algorithm is better known as training algorithm [4], and it has following steps [11]:

1. Initialization of PV parameters.
2. Forward procedure.
3. Backward procedure.

### 3.1.6 Data Validation (DV)

After the PV parameters are determined, we take the symbols from the training algorithm and form an initial sequence of symbols. The recorded sequence is formed up to a particular time instant let us say *t* and is fed in the PV component and compute the probability acceptance. Assuming, the probability is $a_1$. Now the symbol generated by the new data up to time *t+1* and thus forming another sequence. We input this new sequence to the PV component and calculate its new probability acceptance, assuming it to be $a_2$. If the difference between $a_1$ $a_2$ is accepted by the PV component with low probability and thus it prompts for the data invalidation. Otherwise, the data is accepted and is validated by the PV component. Thus the system provides the require results.

### 3.2. Working of DCPS-HMM Model

The task which is fed by the user is fed in the system through the PM (Figure1). It can take several processes at a time and then en-queues all of them in a queue. It keeps the status of the PA and thereby checks its status. If the PA is idle and can take inputs then PM will de-queue the process and pass it to the PA. This de-queued process is en-queued in other list so as to achieve end-to end event scheduling. PA then further characterizes the process/sub process nature and allocates the resources (application processor) to a process/sub process. When the process is passed to the application process then the process is being implemented by the system. If the process as a whole is periodic in nature than it may be given as a whole to the implementation module otherwise it may be segregated into sub modules and is then allocated resources accordingly. Further, it states that aperiodic processes must be further classified according to their need of the resources [12].

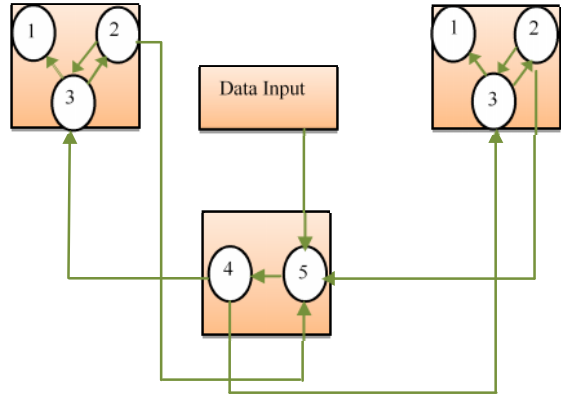

Figure 1.  Distributed Middle Ware Architecture

Table 1.  Notations in Figure 1.

| Component | Notation Discription |
|---|---|
| 1 | Process Validator |
| 2 | Process Tracker |
| 3 | Process Implementation |
| 4 | Processor Allocator |
| 5 | Process Manager |

This system classifies the needs in three categories i.e. *low medium* and *high*. The dimensions for these parameters must be carefully decided at the configuration of the system since it will also decide the validation criteria. But those aperiodic cannot be in *low* range which must be performed in one go.  Thereby, this resource allocator applies the scheduling algorithm (job swapping) to the system. The change in the nature of the process is prompted by the PM leading to change in the scheduling algorithm. Once the process is sent to the application processor, the process implementer does the needful processing and sends the results to the PV which validates the results. If the results are validated then the observation sequence is updated by the latest reading and then the output is displayed.

Meanwhile, the Process Tracker tracks the swapping of periodic process to aperiodic process. If all the sub processes are being processed then that process is removed from the second queue which is formed during passing the new process to the PA.

### 3.3. DCPS-HMM Characteristics

DCPS-HMM has two important characteristics which makes it more feasible i.e.

#### 3.3.1. Job Swapping

It means that some processes/sub processes may be executed in a periodic manner and some of the processes are executed in a periodic manner.

### 3.3.2. State Variation

Means according to the nature of the process the processor will change its mode of implementation. This is very important prospect of a DCPS since it makes the system compatible and flexible with different types of project. It also tests the System's robustness in different environments.

## 4. SIMULATION OF DCPS-HMM ON DIFFERENT SCENARIOS

The Performance of DCPS-HMM model can be simulating using different-different scenarios. Here we take two sample scenarios; one is in credit card fraud detection schemes and second is in CPS-IP. In this section we illustrated both of scenarios in detail using DCPS-HMM model.

### 4.1. Credit Card fraud detection

Credit card purchases can be categorized in two types: 1) Physical card and 2) Virtual card [5]. In a physical card the user presents its card in the front of manufacturer for making a payment. While in case of virtual card only some secret information is being transferred and then the transaction is made. This system provides ample amount of opportunities for the attackers to steal the required information and commit frauds. Therefore, validation of the transaction holds a great deal of importance. Thereby, this proposed DCPS-HMM model provides the solution to these types of frauds by implementing the PV in the system (Figure 3 and 4). Once the user applies for the transaction the transaction details is being en-queued in the list and is passed to the PA by the PM. Since these will be categorized under periodic processes therefore the system can apply scheduling algorithms such as "Round-Robin" in order to improve the efficiency of the system. The PA deploys this data to the implementation module and tracker keeps the track of the process swapping. After, the implementation is completed the result is sent to the PV for checking the transaction consistency.

To map the credit card transaction processing operation in terms of PV we first have to decide the parameters for the validator. In course of this, we assume that the credit card made the transaction to which the DCPS-HMM model takes five states i.e. Very-Low, Low, Medium, High, Very-High. The user when sends the request to the model which when processed by the implementer generates an observation symbol, say *x*. To find the observation symbols corresponding to individual cardholder's transactions dynamically, we run a clustering algorithm on the individual's past transactions generating centroids from the clusters formed [13][14].

The parameters (as in Section 3) T1, T2, w can be decided by the collaboration of an acquiring bank and an issuing bank. However, to make initial data more reliable we take into account the spending profile of customer.

Now, the system will compute the probability of acceptance of transaction before new observation symbol is generated.

$$\alpha_1 = P(O_1, O_2, \ldots, O_R | C) \qquad (1)$$

Let $O_{R+1}$ be the symbol generated by the new credit card transaction. The system now replaces $O_1$ with $O_{R+1}$ and computes its probability of acceptance.

$$\alpha_2 = P(O_2, \ldots, O_{R+1} | C) \qquad (2)$$

If the > 0, the new sequence is accepted by the system with low probability and thus it is rejected.

If the PV approves the transaction, the transaction is committed otherwise the transaction is prompted to be inconsistent.

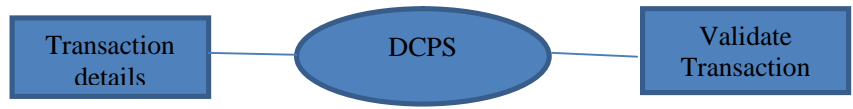

Figure 2. Transaction validation using DCPS

Figure 2 shows the basic steps involves in the process of validation of Transaction details. First transaction details faded into the DCPS model and this model validates the input transactions.

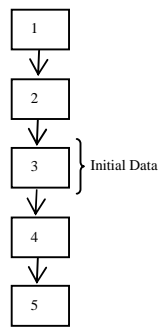
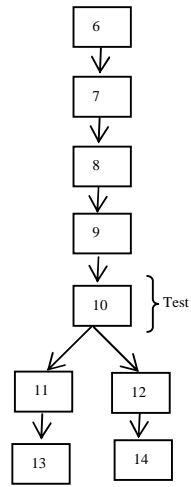

Figure 3. Training

Figure 4. Detection

Table 2. Notations in Figure 2 and Figure 3

| Component | Notation Discription |
|---|---|
| 1 | Create Clusters |
| 2 | Identify the Past Profile |
| 4 | Construct sequence. |
| 5 | Construct the Train Model. |
| 6 | Generate the Observation symbol |
| 7 | Add new observation symbol to generate new sequence |
| 8 | Accept both old and new sequence |
| 9 | Calculate the difference |
| 10 | Test |
| 11 | Normal/Valid |
| 12 | Anomaly/Invalid |
| 13 | Add new observation symbol to the sequence. |
| 14 | Discard the New Observation symbol |

## 4.2. CPS-IP

The transfer of files/data is more frequent in day-to-day life. The IP transfer can be considered more of a two way information exchange in CPS. Therefore, in order to maintain the server's performance assurance one need to be very sure that a particular process (a file exchange for instance) may not send the data in the deadlock state. Further, large chunks may not be given to a processor which is already overloaded is also much of a concern.

The transfer of files through IP will be considered to be an aperiodic process since it will consume the network resources according to the file size. The DCPS-HMM system proposes the solution which is based on the layering concept [12]. It states that the size (parameter for validation) of the file to be transferred will fit in some or the other layer defined. But if this size is not suitable for any of the layers defined then the PV will state the transfer as inconsistent and thus transfer may not take place. If the PA finds that the process is of *low* range aperiodic process then it will apply the job swapping algorithm otherwise it will execute the process in one go. As soon as the process (transfer of file in this case) finishes the transfer it will return the resources back to PA.

Same as the credit card case, the system will form first the clusters in order to find the observation symbols corresponding to the size of the file. We uses K-means clustering algorithm [5] which groups the files of same size range. Thereby, the centroids are generated by the clusters. These centroids are used to decide the observation tables in new sequence. The initial data is chosen by studying the past file transfer record or may be by the network status. Further, the sequence is generated from the training data [5][15] When the user inserts the file for fresh transfer then a new observation symbol is generated adding which we get a new sequence. The new and old symbols are compared which is analyzed to give results. If the PV finds that the file transfer is normal then the file transfer will start otherwise a prompt message for overloading will be displayed.

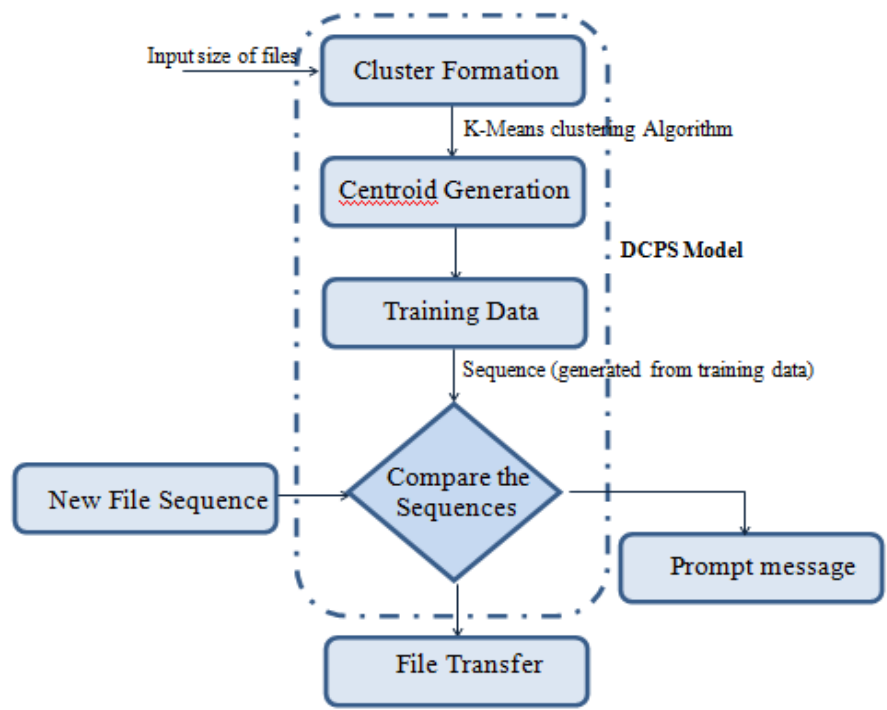

Figure 5. Simulation process of DCPS in CPS-IP File transfer

## 5. CONCLUSION

We conclude that the proposed systems solves the problems faced by many organizations due to false data processed because of the absence of any suitable validation process in distributed systems scenarios. Further, these systems also provide efficient use of resources by having Processor Allocator which focuses on having low degradation of the resources. The proposed system is 1) Robust; as it handles both type of processes (periodic and aperiodic). 2) Secure; as this system proposes a separate component process validator which do not allow any false data to be process by the user as it simulates the past habits of user by using hidden markov model. 3) Efficient; as it has a separate component (processor allocator) which smartly allocates the resources on the nature of process. The proposed DCPS-HMM can be improved in following respects:

1. The Initial configuration for PV of the proposed system is a crucial input for correct validation. Thus, it could be suggested that the proposed system is dependent on the initial data given to the system. Therefore, its dependency can prove to be a hindrance in the correct validation of the result. If this dependency is removed from the approach it will provide better user interface with less system complexity.
2. The Processor Allocation System can be improved in assigning the better scheduling algorithm for big aperiodic process so that the system efficiency is improved significantly.